\documentclass[twocolumn,twoside,slac_two]{revtex4}
\usepackage{graphicx}
\usepackage{epsfig}
\usepackage{fancyhdr}

\usepackage{dcolumn}

\def\reff{\vskip 4pt \par \hangindent 16pt \noindent}

\begin{document}

\title{Optical Design for the Laser Astrometric Test of Relativity}

%

\author{Slava G. Turyshev, Michael Shao}
\affiliation{Jet Propulsion Laboratory, 4800 Oak Grove Drive, Pasadena, CA 91109 USA}
\author{Kenneth L. Nordtvedt, Jr.}
\affiliation{Northwest Analysis, 118 Sourdough Ridge Rd. Bozeman MT 59715 USA}

\begin{abstract}

This paper discusses the Laser Astrometric Test Of Relativity (LATOR) mission. LATOR is a Michelson-Morley-type experiment designed to test the pure tensor metric nature of gravitation - the fundamental postulate of Einstein's theory of general relativity.  With its focus on gravity's action on light propagation it complements other tests which rely on the gravitational dynamics of bodies.  By using a combination of independent time-series of highly accurate gravitational deflection of light in the immediate proximity to the Sun along with measurements of the Shapiro time delay on the interplanetary scales (to a precision respectively better than $10^{-13}$ radians and 1 cm), LATOR will significantly improve our knowledge of relativistic gravity.  The primary mission objective is to i) measure the key post-Newtonian Eddington parameter $\gamma$ with accuracy of a part in 10$^9$.  $(1-\gamma)$ is a direct measure for presence of a new interaction in gravitational theory, and, in its search, LATOR goes a factor 30,000 beyond the present best result, Cassini's 2003 test.  Other mission objectives include: ii) first measurement of gravity's non-linear effects on light to $\sim$0.01\% accuracy; including both the traditional Eddington $\beta$  parameter and also the spatial metric's 2nd order potential contribution (never been measured before);  iii) direct measurement of the solar quadrupole moment $J_2$ (currently unavailable) to accuracy of a part in 200 of its expected size; iv) direct measurement of the ``frame-dragging'' effect on light by the Sun's rotational gravitomagnetic field to one percent accuracy. LATOR's primary measurement pushes to unprecedented accuracy the search for cosmologically relevant scalar-tensor theories of gravity by looking for a remnant scalar field in today's solar system. The key element of LATOR is a geometric redundancy provided by the laser ranging and long-baseline optical interferometry.  We discuss the mission and optical designs  of this proposed experiment. 

\end{abstract}

\maketitle

\thispagestyle{fancy}


\section{Introduction}
\label{sec:intro}

After almost ninety years since general relativity was born, Einstein's theory has survived every test. Such longevity, of course, does not mean that this theory is absolutely correct, but it serves to motivate more accurate tests to determine the level of accuracy at which it is violated. 
Einstein's general theory of relativity (GR) began with its empirical success in 1915 by explaining the anomalous perihelion precession of Mercury's orbit, using no adjustable theoretical parameters.  Shortly thereafter, Eddington's 1919 observations of star lines-of-sight during a solar eclipse confirmed the doubling of the deflection angles predicted by GR as compared to Newtonian-like and Equivalence Principle arguments.  This conformation made the general theory of relativity an instant success. 

From these beginnings, the general theory of relativity has been verified at ever higher accuracy. Thus, microwave ranging to the Viking Lander on Mars yielded accuracy  $\sim$0.2\% in the tests of GR \cite{viking_shapiro1,viking_reasen,viking_shapiro2}. Spacecraft and planetary radar observations reached an accuracy of $\sim$0.15\% \cite{anderson02}.  The astrometric observations of quasars on the solar background performed with Very-Long Baseline Interferometry (VLBI) improved the accuracy of the tests of gravity to $\sim$0.045\% \cite{RoberstonCarter91,Lebach95,Shapiro_SS_etal_2004}. Lunar laser ranging,  a continuing legacy of the Apollo program, provided $\sim$0.011\% verification of GR via precision measurements of the lunar orbit \cite{Ken_LLR68,Ken_LLR91,Ken_LLR30years99,Ken_LLR_PPNprobe03,JimSkipJean96,Williams_etal_2001,Williams_Turyshev_Murphy_2004, LLR_beta_2004}. Finally, the recent experiments with the Cassini spacecraft improved the accuracy of the tests to $\sim$0.0023\% \cite{cassini_ber}. As a result general relativity became the standard theory of gravity when astrometry and spacecraft navigation are concerned. 

However, the tensor-scalar theories of gravity, where the usual general relativity tensor field coexists with one or several long-range scalar fields, are believed to be the most promising extension of the theoretical foundation of modern gravitational theory. The superstring, many-dimensional Kaluza-Klein, and inflationary cosmology theories have revived interest in the so-called `dilaton fields', i.e. neutral scalar fields whose background values determine the strength of the coupling constants in the effective four-dimensional theory. The importance of such theories is that they provide a possible route to the quantization of gravity and unification of physical law. 

Recent theoretical findings suggest that the present agreement between Einstein's theory and experiment might be naturally compatible with the existence of a scalar contribution to gravity. In particular, Damour and Nordtvedt \cite{Damour_Nordtvedt_1993a} (see also 
\cite{DamourPolyakov94} for non-metric versions of this mechanism and \cite{DPV02} for the recent summary of a dilaton-runaway scenario) have found that a scalar-tensor theory of gravity may contain a `built-in' cosmological attractor mechanism towards GR.  A possible scenario for cosmological evolution of the scalar field was given in \cite{Ken_LLR_PPNprobe03,Damour_Nordtvedt_1993a}. Their speculation assumes that the parameter  $\frac{1}{2}(1-\gamma)$  was of order of 1 in the early universe, at the time of inflation, and has evolved to be close to, but not exactly equal to, zero at the present time. In fact, the analyzes discussed above not only motivate new searches for very small deviations of relativistic gravity in the solar system, they also predict that such deviations are currently present in the range from 10$^{-5}$ to $\sim 5\times 10^{-8}$ of the post-Newtonian effects.  This would require measurement of the effects of the next post-Newtonian order ($\propto G^2$) of light deflection resulting from gravity's intrinsic non-linearity. An ability to measure the first order light deflection term at the accuracy comparable with the effects of the second order is of the utmost importance for the gravitational theory and is the challenge for the 21st century fundamental physics. 

When the light deflection in solar gravity is concerned, the magnitude of the first order light deflection effect, as predicted by GR, for the light ray just grazing the limb of the Sun is $\sim1.75$ arcsecond. (Note that 1 arcsecond $\simeq5~\mu$rad; when convenient, below we will use the units of radians and arcseconds interchangeably.) The effect varies inversely with the impact parameter. The second order term is almost six orders of magnitude smaller resulting in  $\sim3.5$ microarcseconds ($\mu$as) light deflection effect, and which falls off inversely as the square of the light ray's impact parameter \cite{Ken_2PPN_87,Turyshev_etal_2004,lator_cqg_2004}. The smallness of the effects emphasize the fact that, among the four forces of nature, gravity is the weakest interaction; it acts at very long distances and controls the large-scale structure of the universe, thus, making the precision tests of gravity a very challenging task. 

This paper discusses the Laser Astrometric Test of Relativity (LATOR)  mission that is designed to directly address the challenges outlined above with an unprecedented accuracy \cite{lator_cqg_2004}. LATOR will test the cosmologically motivated theories that explain the small acceleration rate of the Universe (aka dark energy) via modification of gravity at very large, horizon or super-horizon distances.  This solar system scale experiment would search for a cosmologically-evolved scalar field that is predicted by modern theories of quantum gravity and cosmology, and also by superstring and brane-world models \cite{dvali}.  The value of the Eddington parameter $\gamma$ may be holding a key answer to the most fundamental questions about evolution of the universe.  In the low energy approximation suitable for the solar system, modern theories above predict measurable contributions to the parameter $\gamma$ at the level of $(1-\gamma)\sim 10^{-6}-10^{-8}$; detecting this deviation is the LATOR's primary objective.  With the accuracy of one part in a billion, this mission could discover a violation or extension of general relativity, and/or reveal the presence of any additional long range interaction. 

The paper is organized as follows:  Section \ref{sec:lator} provides the overview for the LATOR experiment including the preliminary mission design. In Section \ref{sec:lator_current} we discuss the current optical design for the LATOR flight system. We also present the expected performance for the LATOR instrument. Section \ref{sec:conc} discusses the next steps that will be taken in the development of the LATOR mission.  

\section{The LATOR Mission}
\label{sec:lator}

\begin{figure*}[t!]
 \begin{center}
\noindent    
\psfig{figure=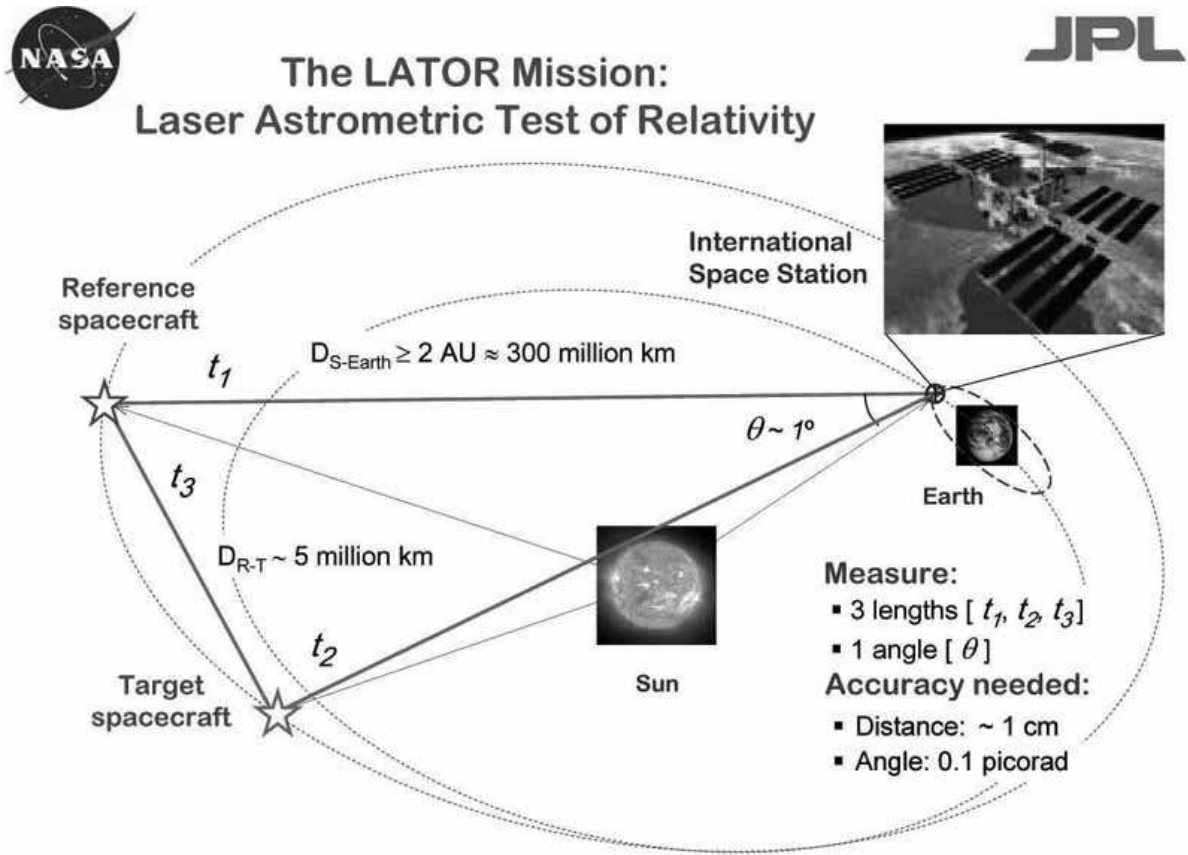,width=158mm}
\end{center}
\vskip -10pt 
  \caption{The overall geometry of the LATOR experiment.  
 \label{fig:lator}}
\end{figure*} 


The LATOR mission architecture uses an evolving light triangle formed by laser ranging between two spacecraft (placed in $\sim$1 AU heliocentric orbits) and a laser transceiver terminal on the International Space Station (ISS: via European collaboration).  The objective is to measure gravitational deflection of laser light as it passes in extreme proximity to the Sun (see Figure \ref{fig:lator}).  To that extent, the long-baseline ($\sim$100 m) fiber-coupled optical interferometer on the ISS will perform differential astrometric measurements of the laser light sources on the two spacecraft as their lines-of-sight pass behind the Sun.  As seen from the Earth, the two spacecraft will be separated by about 1$^\circ$, which will be accomplished by a small maneuver immediately after their launch \cite{lator_cqg_2004}. This separation would permit differential astrometric observations to accuracy of $\sim 10^{-13}$ radians needed to significantly improve measurements of  gravitational deflection of light by the solar gravity.

To enable the primary objective, LATOR will place two spacecraft into a heliocentric orbit so that observations may be made when the spacecraft are behind the Sun as viewed from the ISS.  To avoid having to make absolute measurements, the spacecraft will be placed in a 3:2 Earth resonant orbit that provides three observing sessions during the initial 21 months after the launch with the first session starting in 15 months \cite{lator_cqg_2004}. Such an orbit provides significant variation of the distance between the beam and the center of the Sun (i.e impact parameter); the parameters will vary from 10 to 1 solar radii over a period of $\sim$20 days. The three arms of the triangle will be monitored with laser ranging, based on the time-of-flight measurements and will be accurate to $\sim 1$ cm. From three measurements one can calculate the Euclidean value for any angle in this triangle.  

As evident from Figure \ref{fig:lator}, the key element of the LATOR experiment is a redundant geometry optical truss to measure departure from Euclidean geometry ($\sim 8\times 10^{-6}$) caused by the solar gravity field.  This departure is shown as a difference between the calculated Euclidean value for an angle in the triangle and its value directly measured by the interferometer.  The discrepancy is due to the curvature of the space-time around the Sun, it is computed for every alternative theory of gravity and it constitutes the LATOR's signal of interest.  The built-in redundancy eliminates the need for drag-free spacecraft for high-accuracy navigation \cite{lator_cqg_2004}. 
Therefore, the uniqueness of this mission comes with its built-in geometrically redundant architecture that enables LATOR to measure the departure from Euclidean geometry caused by the solar gravity field to a very high accuracy. The accurate measurement of this departure constitutes the primary mission objective.

\subsection{Science with LATOR}

\begin{table*}[ht!]
\caption{LATOR Mission Summary: Science Objectives 
\label{tab:summ_science}}
\vskip 5pt
\begin{tabular}{m{12.0cm}} \hline \hline 

\reff\hskip4pt$\bullet$\hskip6pt
To test Einstein's general theory of relativity in the most intense gravitational environment available in the solar system -- the extreme proximity to the Sun

\reff\hskip4pt$\bullet$\hskip6pt
To measure the key Eddington PPN parameter $\gamma$ with accuracy of 1 part in 10$^{9}$ -- a factor of 30,000 improvement in the tests of gravitational deflection of light

\reff\hskip4pt$\bullet$\hskip6pt
To provide direct and independent measurement of the Eddington PPN parameter $\beta$ via gravity effect on light to $\sim0.01$\% accuracy

\reff\hskip4pt$\bullet$\hskip6pt
To measure effect of the 2-nd order gravitational deflection of light with accuracy of $\sim1\times 10^{-4}$, including first ever measurement of the PPN parameter $\delta$ 

\reff\hskip4pt$\bullet$\hskip6pt
To measure the solar quadrupole moment $J_2$ (using the theoretical value of the solar quadrupole moment $J_2\simeq10^{-7}$) to 1 part in 200, currently unavailable

\reff\hskip4pt$\bullet$\hskip6pt
To directly measure the frame dragging effect on light (first such observation) with $\sim 1\times10^{-3}$ accuracy 


\reff\hskip4pt$\bullet$\hskip6pt
To test alternative theories of gravity and cosmology (i.e scalar-tensor theories) by searching for cosmological remnants of scalar field in the solar system
\\[2pt]\hline 
\end{tabular}
\end{table*}
 

LATOR is a Michelson-Morley-type experiment designed to test the pure tensor metric nature of gravitation - the fundamental postulate of Einstein's theory of general relativity \cite{lator_cqg_2004}.  With its focus on gravity's action on light propagation it complements other tests which rely on the gravitational dynamics of bodies.  By using a combination of independent time-series of highly accurate gravitational deflection of light in the immediate proximity to the Sun along with measurements of the Shapiro time delay on the interplanetary scales (to a precision respectively better than $10^{-13}$ radians and 1 cm), LATOR will significantly improve our knowledge of relativistic gravity.  

The primary mission objective is to measure the key post-Newtonian Eddington parameter $\gamma$ with accuracy of a part in 10$^9$.  This parameter, whose value in GR is unity, is perhaps the most fundamental PPN parameter, in that $(1-\gamma)$ is a direct measure for presence of a new interaction in gravitational theory \cite{Damour_Nordtvedt_1993a,Damour_EFarese96,lator_cqg_2004}. Within perturbation theory for such theories, all other PPN parameters to all relativistic orders collapse to their GR values in proportion to $(1-\gamma)$. This is why measurement of the first order light deflection effect at the level of accuracy comparable with the second-order contribution would provide the crucial information separating alternative scalar-tensor theories of gravity from the general theory of relativity \cite{Ken_2PPN_87} and also to probe possible ways for gravity quantization and to test modern theories of cosmological evolution \cite{Damour_Nordtvedt_1993a,DamourPolyakov94,DPV02,dvali}  discussed in the previous section.  LATOR is designed to directly address this issue with an unprecedented accuracy and in its search, LATOR goes a factor 30,000 beyond the present best result, Cassini's 2003 test \cite{cassini_ber,luciano}. It will also reach ability to measure the next post-Newtonian order ($\propto G^2$) of light deflection with accuracy to 1 part in $10^3$.

Other mission objectives include: ii) first measurement of gravity's non-linear effects on light to $\sim$0.01\% accuracy; including both the traditional Eddington   parameter and also the spatial metric's 2nd order potential contribution (never been measured before);  iii) direct measurement of the solar quadrupole moment $J_2$ (currently unavailable) to accuracy of a part in 200 of its expected size; iv) measuring the direct ``frame-dragging'' effect on light by the Sun's rotational gravitomagnetic field to one percent accuracy. LATOR's primary measurement pushes to unprecedented accuracy the search for cosmologically relevant scalar-tensor theories of gravity by looking for a remnant scalar field in today's solar system. 

The goal of measuring deflection of light in solar gravity with accuracy of one part in $10^{8}$ requires serious consideration of systematic errors. This work requires a significant effort to properly identify the entire set of factors that may influence the accuracy at this level. Fortunately, we initiated this process being aided with experience of developing a number of instruments that require similar technology and a comparable level of accuracy, notably SIM, TPF, Keck and Palomar Testbed Interferometers. This experience comes with understanding various constituents of the error budget, expertise in developing appropriate instrument models; it is also supported by the extensive verification of the expected  performance with the set of instrumental test-beds and existing flight hardware. Details of the LATOR error budget are still being developed and will be published elsewhere, when fully analyzed. 

Here we discuss a the LATOR astrometric observable as it relates to the realization of the future optical infrastructure.

\subsection{Observational Model for LATOR} 

In development of the mission's error budget we use a simple model to capture all error sources and their individual impact on the mission performance \cite{Turyshev_etal_2004}. The light paths, $\ell_{ij}$, between the three vortices of the triangle  may be given by an expression for the Shapiro time delay relation, that to the first order in gravitational constant, has the form: 
\begin{equation}
\ell_{ij}=r_{ij}+(1+\gamma)\mu_\odot\ln[\frac{r_i+r_j+r_{ij}}{r_i+r_j-r_{ij}}], ~~~~ \mathbf{r}_{ij}=\mathbf{r}_j-\mathbf{r}_i,
\label{eq:path}
\end{equation}
\noindent where $\mathbf{r}_i$ is the barycentric Euclidian position to one of the three vortices, $i,j\in\{1,3\}$ ($i=3$ is for the ISS), with $r_i=|\mathbf{r}_i|$, being its distance, and $\mu_\odot=GM/c^2$ is the solar gravitational radius. To a similar accuracy, the interferometric delay, $ d_{j}$, for a laser source $j$ has the following approximate form (i.e. differenced Shapiro time delay for the two telescopes separated by an interferometric baseline, $\mathbf{b}$, or $d_j=\ell_{j3}(\mathbf{r}_3)-\ell_{j3}(\mathbf{r}_3+\mathbf{b})$):
{}
\begin{equation}
d_j\simeq(\mathbf{b}\cdot\mathbf{n}_{j3})-(1+\gamma)\mu_\odot 
\frac{2r_jr_3}{r_3+r_j}\frac{\mathbf{b}\cdot(\mathbf{n}_3-\mathbf{n}_{j3})}{p_j^2},
\label{eq:delay}
\end{equation} 
\noindent where $p_j$ is the solar impact parameter for source $j$. Both expressions Eqs.(\ref{eq:path}) and (\ref{eq:delay})  require some additional transformations to keep only the terms with a similar order. The entire LATOR model accounts for a whole range of other effects, including due to gravitational multipoles, second order deflection, angular momentum contribution, and etc. This work had being initiated and the corresponding results will be reported elsewhere. Below we comment only on the conceptual formulation of the LATOR observables. 

The range observations Eq.(\ref{eq:path}) may be used to measure any angle between the three fiducials in the triangle. However, for observations in the solar gravity field, measuring the lengths do not give you a complete information to determine the angles, and some extra information is needed. This information is the mass of the Sun, and, at least one of the impact parameters. Nevertheless, noting that the paths $\underline{\ell}_{ij}$ correspond to the sides of the connected, but gravitationally distorted triangle, one can write $\underline{\ell}_{12}+\underline{\ell}_{23}+\underline{\ell}_{31}=0$, where $\underline{\ell}_{ij}$ is the null geodesic path for light moving between the two points $i$ and $j$. This leads to the expression for the angle between the spacecraft $\cos(\widehat{\underline{\ell}_{31}\underline{\ell}_{32}})=\cos\delta_r =(\ell_{32}^2+\ell_{31}^2-\ell_{12}^2)/(2 \ell_{32} \ell_{31})$. Expression for $\cos\delta_r$ will have both Euclidian and gravitational contributions; their detailed form will not significantly contribute to the discussion below and, thus, it is outside the scope of this paper.

The astrometric observations Eq.(\ref{eq:delay}) will be used to obtain another measurement of the same angle between the two spacecraft. The LATOR interferometer will perform differential observations between the two sources of laser light, measuring the differential delay $\Delta d_{12}=d_2-d_1$ to the accuracy of less than $5$ pm (see below). For the appropriate choice of the baseline orientation, one can present the angle between the two sources of laser light as  $\cos(\widehat{\underline{\ell}_{31}\underline{\ell}_{32}})=\cos\delta_a=1-\Delta d_{12}^2/2b^2$. This expression would have both Euclidian and gravity contributions which are not discussed in detail in this paper.

The two sets of observations obtained by laser ranging and astrometric interferometry form the complete set of LATOR observables. Conceptually,  the LATOR astrometric measurement $\delta_d$ of the gravitational deflection of light may be modeled as
\begin{equation}
\delta_d = \delta_r-\delta_a = c_1\Big(\frac{1}{p} - \frac{1}{p+\Delta p}\Big)+c_2\Big(\frac{1}{p^2} - \frac{1}{(p+\Delta p)^2}\Big),
\end{equation}
 
\noindent where  $\delta_r$ is the angle computed from the range information,  $\delta_a$ is the angle measured astrometrically by the interferometer. $p$ is the impact parameter of the spacecraft closer to the Sun and $\Delta p$ is the difference between the two impact parameters. $c_1$ and $c_2$ are the first and second order terms in the gravitational deflection and are the quantities of interest. Three such measurements are made to simultaneously solve for these constants together with the impact parameter. 
The temporal evolution of the entire triangle structure will produce another set of observables -- the range rate data, expressed as $\delta_v=d\delta_d/dt=(\partial \delta_d/\partial p)dp/dt$ which will also be used to process the data. A fully relativistic model for this additional independent observable, including the contributions of range and angle rates, is being currently developed \cite{Turyshev_etal_2004}.  The error budget is subdivided into three components -- range and interferometer measurements, and spacecraft stability that all relate to the expected performance of the optical system. 

As evident from Figure \ref{fig:lator}, the key element of the LATOR experiment is a redundant geometry optical truss to measure the effects of gravity on the laser signal trajectories.  LATOR will generate four time series of measurements -- one for the optical range of each side of the triangle, plus the angle between light signals arriving at one vertex of the light triangle.  Within the context of a moving Euclidean light triangle, these measurements are redundant.  From a combination of these four times series of data, the several effects of gravity on the light propagations can be precisely and separately determined.  For example: the first and second order gravity monopole deflections go as $1/p$ and $1/p^2$ while the solar quadrupole deflection goes as $1/p^3$ [with $p(t)$ being a laser signal's evolving impact parameter]; the quadrupole moment's deflection has further latitude dependence if spacecraft lines of sight are so located.  The data will be taken over periods in which the laser light's impact parameters $p(t)$ vary from one to ten solar radii, producing time signatures in the data which permits both the separation of the several gravitational effects and the determination of key spacecraft location coordinates needed to calibrate the deflection signals.  The temporal evolution of the entire triangle structure will produce the range rate and angle rate data that will be used to process the experimental data. 

We shall now consider the basic elements of the LATOR optical design. 

\section{Optical Design}
\label{sec:lator_current}

\begin{figure*}[t!]
 \begin{center}
\noindent    
\psfig{figure=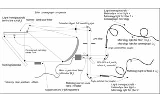,width=155mm}
\end{center}
\vskip -10pt 
  \caption{Basic elements of optical design for the LATOR interferometer: The laser light (together with the solar background) is going through a full aperture ($\sim20$cm) narrow band-pass filter with $\sim 10^{-4}$ suppression properties. The remaining light illuminates the baseline metrology corner cube and falls onto a steering flat mirror where it is reflected to an off-axis telescope with no central obscuration (needed for metrology). It is then enters the solar coronograph compressor by first going through a 1/2 plane focal plane occulter and then coming to a Lyot stop. At the Lyot stop, the background solar light is reduced by a factor of $10^{6}$. The combination of a narrow band-pass filter and coronograph enables the solar luminosity reduction from $V=-26$ to $V=4$ (as measured at the ISS), thus, enabling the LATOR precision observations.
\label{fig:optical_design}}
\end{figure*}

A single aperture of the interferometer on the ISS consists of three 20 cm diameter telescopes (see Figure \ref{fig:optical_design} for a conceptual design). One of the telescopes with a very narrow bandwidth laser line filter in front and with an InGAs camera at its focal plane, sensitive to the 1064 nm laser light, serves as the acquisition telescope to locate the spacecraft near the Sun.

The second telescope emits the directing beacon to the spacecraft. Both spacecraft are served out of one telescope by a pair of piezo controlled mirrors placed on the focal plane. The properly collimated laser light ($\sim$10W) is injected into the telescope focal plane and deflected in the right direction by the piezo-actuated mirrors. 

The third telescope is the laser light tracking interferometer input aperture which can track both spacecraft at the same time. To eliminate beam walk on the critical elements of this telescope, two piezo-electric X-Y-Z stages are used to move two single-mode fiber tips on a spherical surface while maintaining focus and beam position on the fibers and other optics. Dithering at a few Hz is used to make the alignment to the fibers and the subsequent tracking of the two spacecraft completely automatic. The interferometric tracking telescopes are coupled together by a network of single-mode fibers whose relative length changes are measured internally by a heterodyne metrology system to an accuracy of less than 10 pm.

The spacecraft  are identical in construction and contain a relatively high powered (1 W), stable (2 MHz per hour $\sim$  500 Hz per second), small cavity fiber-amplified laser at 1064 nm. Three quarters of the power of this laser is pointed to the Earth through a 10 cm aperture telescope and its phase is tracked by the interferometer. With the available power and the beam divergence, there are enough photons to track the slowly drifting phase of the laser light. The remaining part of the laser power is diverted to another telescope, which points towards the other spacecraft. In addition to the two transmitting telescopes, each spacecraft has two receiving telescopes.  The receiving telescope, which points towards the area near the Sun, has laser line filters and a simple knife-edge coronagraph to suppress the Sun light to 1 part in $10^4$ of the light level of the light received from the space station. The receiving telescope that points to the other spacecraft is free of the Sun light filter and the coronagraph.

In addition to the four telescopes they carry, the spacecraft also carry a tiny (2.5 cm) telescope with a CCD camera. This telescope is used to initially point the spacecraft directly towards the Sun so that their signal may be seen at the space station. One more of these small telescopes may also be installed at right angles to the first one to determine the spacecraft attitude using known, bright stars. The receiving telescope looking towards the other spacecraft may be used for this purpose part of the time, reducing hardware complexity. Star trackers with this construction have been demonstrated many years ago and they are readily available. A small RF transponder with an omni-directional antenna is also included in the instrument package to track the spacecraft while they are on their way to assume the orbital position needed for the experiment. 

In the next Section we present elements for the LATOR optical receiver system.  While we focus on the optics for the two spacecraft, the interferometer has essentially similar optical architecture. 


\begin{table*}[t!]
\begin{center}
\caption{Summary of design parameters for the LATOR optical receiver system.
\label{table:requirements}} \vskip 8pt
\begin{tabular}{rl} \hline\hline
Parameters/Requirements   & Value/Description \\\hline
 & \\[-10pt]
Aperture &  100 mm, unobstructed \\[3pt]
Wavelength & 1064 nm \\[3pt]
Narrow bandpass Filter & 2 nm FWHM over full aperture \\[3pt] 
Focal Planes & APD Data \& CCD Acquisition/Tracking \\[3pt]
APD Field of View & Airy disk field stop (pinhole) in front of APD\\[3pt]
APD Field Stop (pinhole) & Approximately 0.009 mm in diameter \\[3pt] 
APD Detector Size & TBD (a little larger than 0.009 mm)\\[3pt] 
CCD Field of View & 5 arc minutes \\[3pt] 
CCD Detector Size & 640 $\times$ 480 pixels (9.6 mm $\times$ 7.2 mm)\\[3pt]
CCD Detector Pixel Size & 15 $\mu$m\\[3pt] 
Beamsplitter Ratio (APD/CCD) & 90/10\\[3pt] 
Field Stop & `D'-shaped at primary mirror focus \\[3pt] 
Lyot Stop & Circular aperture located at telescope exit pupil\\[3pt] 
\hline\hline
\end{tabular} 
\end{center} 
\end{table*}

\begin{figure*}[t!]
 \begin{center}
\noindent    
\psfig{figure=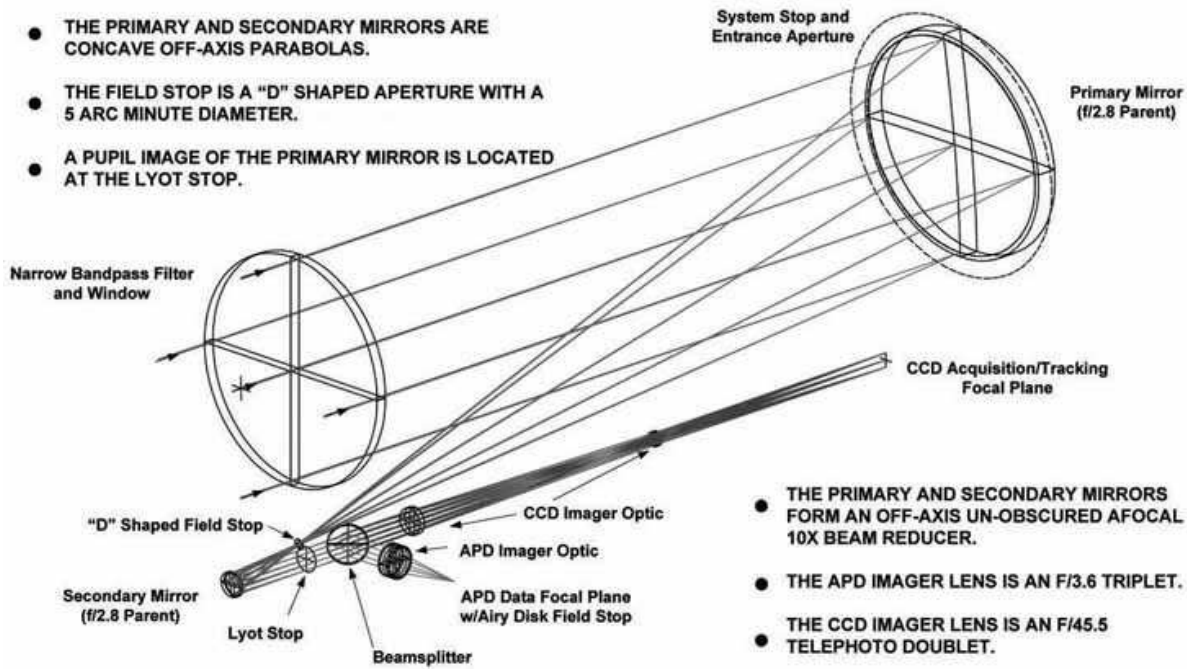,width=158mm}
\end{center}
\vskip -10pt 
  \caption{LATOR receiver optical system layout.  
 \label{fig:lator_receiver}}
\end{figure*} 

\begin{figure*}[t!]
 \begin{center}
\noindent    
\psfig{figure=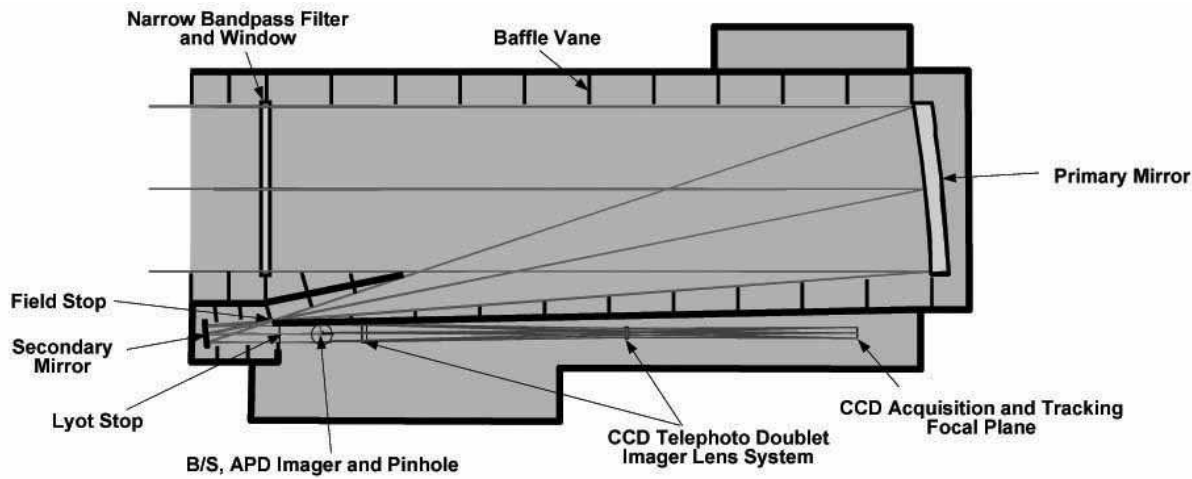,width=160mm}
\end{center}
\vskip -10pt 
  \caption{The LATOR preliminary baffle design.  
 \label{fig:lator_buffle}}
\end{figure*} 
\begin{figure*}[t!]
 \begin{center}
\noindent    
\psfig{figure=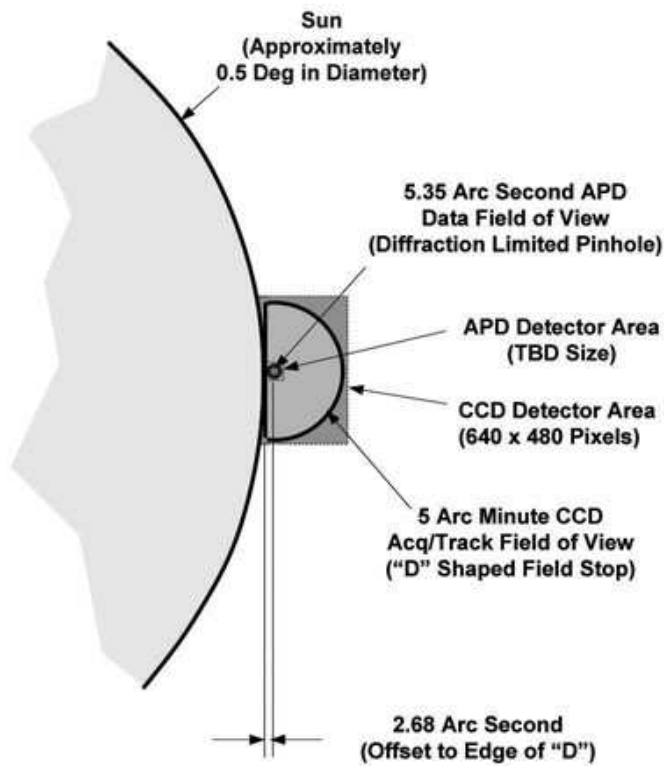,width=92mm}
\end{center}
\vskip -10pt 
  \caption{LATOR focal plane mapping (the diagram not to scale).  
 \label{fig:lator_focal}}
\end{figure*} 

\subsection{The LATOR Optical Receiver System}

The LATOR 100 mm receiver optical system is a part of a proposed experiment. This system is located at each of two separate spacecraft placed on heliocentric orbits, as shown in Figure \ref{fig:lator}. The receiver optical system receives optical communication signals form a transmitter on the ISS, that is in orbit around the Earth. To support the primary mission objective, this system must be able to receive the optical communication signal from the uplink system at the ISS that passes through the solar corona at the immediate proximity of the solar limb (at the distance no more then 5 Airy disks). 

Our recent analysis of the LATOR 100 mm receiver optical system successfully satisfied all the configuration and performance requirements (shown in Table \ref{table:requirements}). We have also performed a conceptual design (see Figure \ref{fig:lator_receiver}), which was validated with a CODEV  ray-trace analysis. The ray-trace performance of the designed instrument is diffraction limited in both the APD and CCD channels over the specified field of view at 1064 nm. The design incorporated the required field stop and Layot stop. A preliminary baffle design has been developed for controlling the stray light.

The optical housing is estimated to have very accommodating dimensions; it measures (500 mm $\times$ 150 mm $\times$ 250 mm). The housing could be made even shorter by reducing the focal length of the primary and secondary mirrors, which may impose some fabrication difficulties. These design opportunities are being currently investigated. 

\subsubsection{Preliminary Baffle Design}

Figure \ref{fig:lator_buffle} shows the LATOR preliminary baffle design. The out-of-field solar radiation falls on the narrow band pass filter and primary mirror; the scattering from these optical surfaces puts some solar radiation into the FOV of the two focal planes. This imposes some requirements on the instrument design.  Thus, the narrow band pass filter and primary mirror optical surfaces must be optically smooth to minimize narrow angle scattering. This may be difficult for the relatively steep parabolic aspheric primary mirror surface. However, the field stop will eliminate direct out-of-field solar radiation at the two focal planes, but it will not eliminate narrow angle scattering for the filter and primary mirror.  Finally, the Lyot stop will eliminate out-of-field diffracted solar radiation at the two focal planes. Additional baffle vanes may be needed several places in the optical system. This design will be further investigated in series of trade-off studies with support from this proposal by also focusing on the issue of stray light analysis. 

\begin{figure*}[t!]
 \begin{center}
\noindent    
\psfig{figure=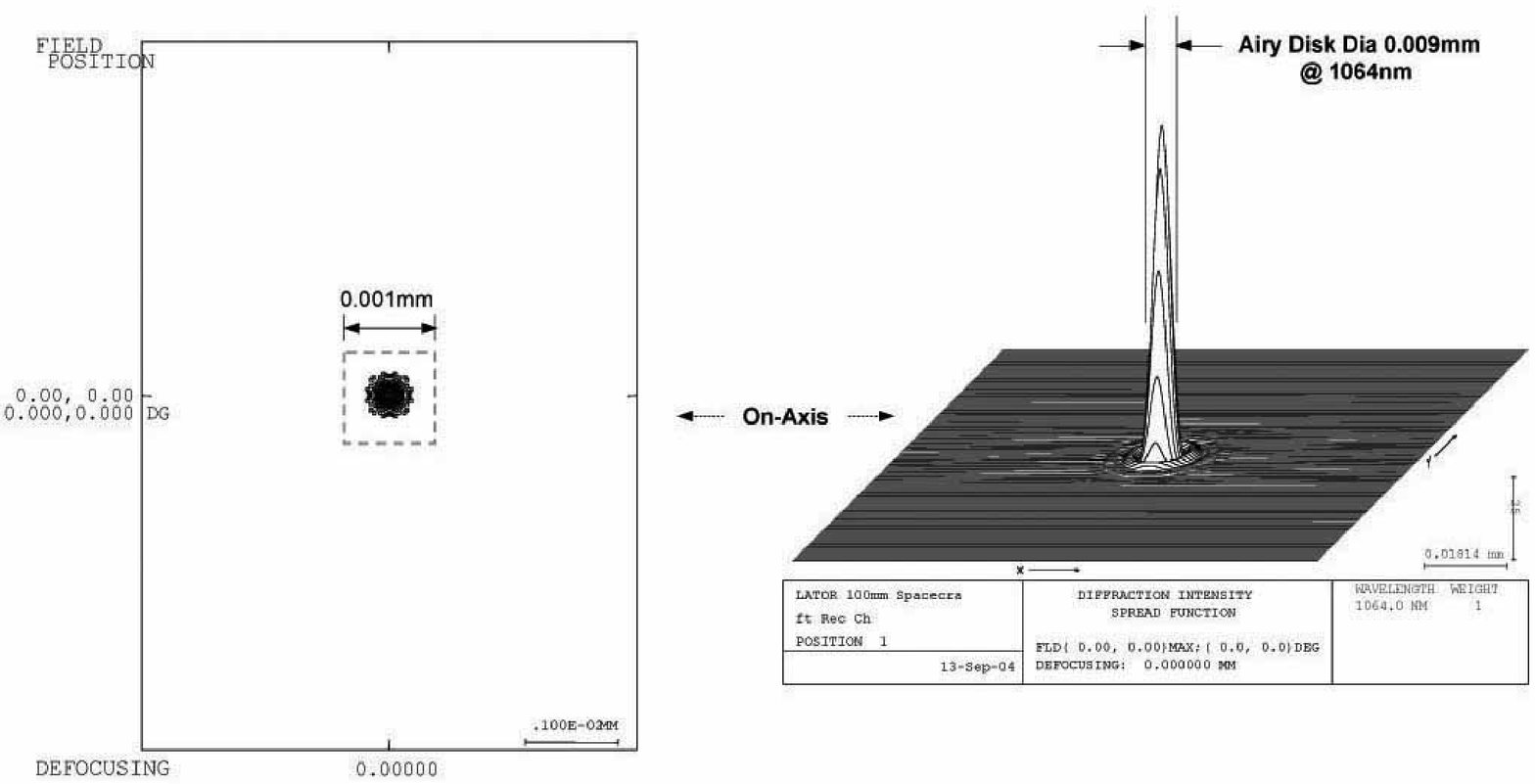,width=158mm}
\end{center}
\vskip -10pt 
  \caption{APD channel geometric (left) and diffraction (right) PSF.  
 \label{fig:lator_apd}}
 \begin{center}
\noindent    
\psfig{figure=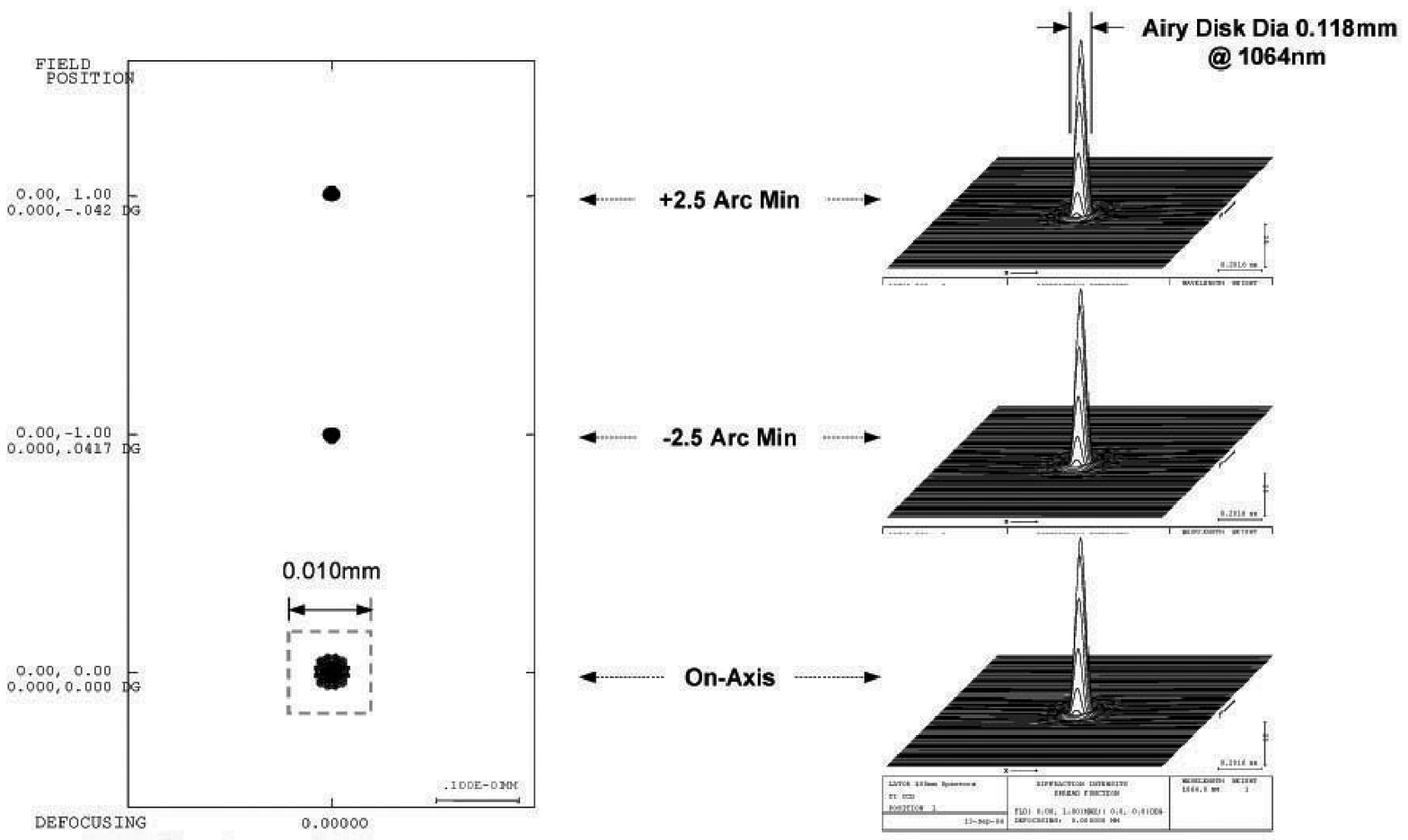,width=158mm}
\end{center}
\vskip -10pt 
  \caption{CCD channel geometric (left) and diffraction (right) PSF.  
 \label{fig:lator_ccd}}
\end{figure*} 

\subsubsection{Focal Plane Mapping}

Figure \ref{fig:lator_focal} shows the design of the focal plane capping. The straight edge of the `D'-shaped CCD field stop is tangent to the limb of the Sun and it is also tangent to the edge of APD field stop (pinhole). There is a 2.68 arcsecond offset between the straight edge and the concentric point for the circular edge of the CCD field stop (`D'-shaped aperture). In addition, the APD field of view and the CCD field of view circular edges are concentric with each other. Depending on the spacecraft orientation and pointing ability, the `D'-shaped CCD field stop aperture may need to be able to be rotated to bring the straight edge into a tangent position relative to the limb of the Sun. 
The results of the analysis of APD and CCD channels point spread functions (PSF) are shown in Figures \ref{fig:lator_apd} and \ref{fig:lator_ccd}.

\subsection{Factors Affecting SNR Analysis}

In conducting the signal-to-noise analysis we pay significant attention to several important factors. In particular, we estimate what fraction of the transmitted signal power captured by the 100 mm receiver aperture and analyze the effect of the Gaussian beam divergence (estimated at $\sim 7 ~\mu$rad) of the 200 mm transmit aperture on the ISS. Given the fact that the distance between the transmitter and receiver is on the order of 2 AU, the amount captured is about $2.3\times 10^{-10}$ of the transmitted power.  

We also consider the amount of solar disk radiation scattered into the two receiver focal planes. In particular, the surface contamination, coating defects, optical roughness and substrate defects could scatter as much as $1\times10^{-4}$ or more (possibly $1\times 10^{-3}$)  of the solar energy incident on the receive aperture into the field of view.   These issues are being considered in our current analysis. We also study the amount of the solar corona spectrum within the receive field of view that is not blocked by the narrow band pass filter.  The factors we consider is the filter's FWHM band-pass is 2 nm, the filter will have 4.0 optical density (OD) blocking outside the 2 nm filter band pass from the X-ray region of 1200 nm; the filter efficiency within the band pass will be about 35\%, and the detector is probably sensitive from 300 nm to 1200 nm.  

Additionally, we consider the amount of out-of-field solar radiation scattered into the focal plane by the optical housing. This issue needs to be investigated in a stray light analysis which can be used to optimize the baffle design to minimize the stray light at the focal plane.  Finally, we study the effectiveness of the baffle design in suppressing stray light at the focal plane. Thus, in addition to the stray light analysis, the effectiveness of the final baffle design should be verified by building an engineering model that can be tested for stray light.  


The importance of this design is in the fact that it can be applied for many applications, thus, opening new ways for optical communication, accuracy navigational and fundamental physics experiments. This LATOR-related design experience motivates us to think about an architecture that may have a much border uses for the purposes of precision navigation and high data rate transmission and capable to operate at large interplanetary distance. In the next section we will summarize our current ideas.

\section{Conclusions}
\label{sec:conc}

The LATOR experiment benefits from a number of advantages over techniques that use radio waves to study the light propagation in the solar vicinity.  The use of monochromatic light enables the observation of the spacecraft almost at the limb of the Sun, as seen from the ISS.  The use of narrowband filters, coronagraph optics, and heterodyne detection will suppress background light to a level where the solar background is no longer the dominant noise source.  The short wavelength allows much more efficient links with smaller apertures, thereby eliminating the need for a deployable antenna.  Advances in optical communications technology allow low bandwidth telecommunications with the LATOR spacecraft without having to deploy high gain radio antennae needed to communicate through the solar corona.  Finally, the use of the ISS not only makes the test affordable, but also allows conducting the experiment above the Earth's atmosphere   the major source of astrometric noise for any ground based interferometer.  This fact justifies the placement of LATOR's interferometer node in space. 

The experiment will utilize several technology solutions that recently became available. In particular, signal acquisition on the solar background will be done with a full-aperture narrow band-pass filer and coronagraph. The issue of the extended structure vibrations of the will be addressed by using $\mu$-g accelerometers. (The use of the accelerometers was first devised for SIM, but at the end their utilization is not needed. The Keck Interferometer uses accelerometers extensively.) Finally, the problem of monochromatic fringe ambiguity that complicated the design of the previous version of the experiment \cite{lator_cqg_2004} and led to the use of variable baselines lengths -- is not an issue for LATOR. This is because the orbital motion of the ISS provides variable baseline projection that eliminates this problem for LATOR.  

The concept is technologically sound; the required technologies have been demonstrated as part of the Space Interferometry Mission developments at JPL.   The LATOR experiment does not need a drag-free system, but uses a geometric redundant optical truss to achieve a very precise determination of the interplanetary distances between the two micro-spacecraft and a beacon station on the ISS. The interest of the approach is to take advantage of the existing space-qualified optical technologies leading to an outstanding performance in a reasonable mission development time. In addition, the issues of the extended structure vibrations on the ISS, interferometric fringe ambiguity, and signal acquisition on the solar backgrounds have all been analyzed, and do not compromise mission goals.  The ISS is the default location for the interferometer, however, ground- and free-flying versions have also been studied.  While offering programmatic benefits, these options differ in cost, reliability and performance. The  availability of the ISS (via European collaboration) makes this mission concept realizable in the very near future. A recent JPL Team X study confirmed the feasibility of LATOR as a NASA Medium Explorer (MIDEX) class mission. 

LATOR is envisaged as a partnership between NASA and ESA wherein both partners are essentially equal contributors, while focusing on different mission elements: NASA provides the deep space mission components and interferometer design, while building infrastructure on the ISS is an ESA contribution. The NASA focus is on mission management, system engineering, software management, integration (both of the payload and the mission), the launch vehicle for the deep space component, and operations. The European focus is on interferometer components, the initial payload integration, optical assemblies, testing of the optics in a realistic ISS environment. In their recent decision, the ESA Panel on the Physical Sciences of the ESA Directorate for Human Space Flight, Microgravity and Exploration supported LATOR as one of their focus missions.  This decision opens for LATOR direct access to the ISS.  The proposed arrangement would provide clean interfaces between familiar mission elements.

This mission may become a 21st century version of Michelson-Morley experiment in the search for a  cosmologically evolved scalar field in the solar system. As such, LATOR will lead to very robust advances in the tests of fundamental physics: it could discover a violation or extension of general relativity, or reveal the presence of an additional long range interaction in the physical law.  There are no analogs to the LATOR experiment; it is unique and is a natural culmination of solar system gravity experiments.

\bigskip 
\begin{acknowledgments}
The work described here was carried out at the Jet Propulsion Laboratory, California Institute of Technology, under a contract with the National Aeronautics and Space Administration.
\end{acknowledgments}

\bigskip 

\end{document}